\documentstyle[12pt,epsf]{article}

\newcommand{\vs}[1]{\vspace*{#1}}
\newcommand{\hs}[1]{\hspace*{#1}}
\newcommand{\del}{\partial}

\begin{document}

\begin{titlepage}

\title{
\vspace{-10mm}
\hfill\parbox{4cm}{
{\normalsize KEK-TH-780}\\[-5mm]
{\normalsize UT-960}\\[-5mm]
{\normalsize\tt hep-th/0108109}
}
\\
\vspace{15mm}
Simple Brane World Scenario with Positive Five Dimensional Cosmological
 Constant
}
\author{
{}
\\
Shoko {\sc Hayakawa}${}^1$\thanks{{\tt shoko@hep-th.phys.s.u-tokyo.ac.jp}},
\hs{1mm}
Takayuki {\sc Hirayama}${}^2$\thanks{{\tt thirayam@post.kek.jp}}
\hs{1mm}
and
\hs{1mm}
Ryuichiro {\sc Kitano}${}^{2, 3}$\thanks{{\tt ryuichiro.kitano@kek.jp}}
\\[15pt]
${}^1${\it Department of Physics, University of Tokyo,}\\
{\it Tokyo, 113-0033, Japan}\\[7pt]
${}^2${\it High Energy Accelerator Research Organization (KEK),}\\
{\it Tsukuba, Ibaraki 305-0801, Japan}\\[7pt]
${}^3${\it Department of Particle and Nuclear Physics,}\\
{\it The Graduate University for Advanced Studies,}\\
{\it Tsukuba, Ibaraki 305-0801, Japan}\\[7pt]
}
\date{\normalsize August, 2001}
\maketitle
\thispagestyle{empty}

\begin{abstract}
\normalsize\noindent
We present a simple brane-world model in five dimensions. In this
model we do not need any fine-tuning between the five dimensional 
cosmological constant and the brane tension to obtain 
four dimensional flat Minkowski space. 
The space-time of our solution has no naked singularities.
Further the compactification scale of the fifth direction is 
automatically determined.
\end{abstract}

\end{titlepage}

\section{Introduction}
The idea we are living on a brane in a higher dimensional bulk 
has been considered since early 1980's \cite{rub1,aka}. After
the discovery of D-brane in string theories on which some Yang-Mills
fields and matter fields are localized, brane world scenario has been
intensely discussed.
Among various models, Randall and Sundrum gave simple ones 
in a five dimensional space-time 
with a negative bulk cosmological constant \cite{rs1,rs2}. In paper
\cite{rs1} they realized a large hierarchy using exponential warp
factor, while in \cite{rs2}, they realized four dimensional gravity 
on a brane even in an infinitely large extra dimension. 
Moreover brane world scenario opens the possibility to
approach the cosmological constant problem. Higher-dimensional approach
to this problem was considered in paper \cite{rub}. Recently
the authors of \cite{ham, kac} showed flat four dimensional space time
can be realized regardless of the Standard Model contribution to the
cosmological constant and this mechanism is sometimes called self-tuning
mechanism. These models have changed dramatically the
commonly assumed properties of Kaluza-Klein models.

There are much interesting points in brane world scenario, but some
difficulties still remain. In the original Randall-Sundrum models
\cite{rs1,rs2} brane tensions must be fine-tuned in order to
get a flat four dimensional space-time. 
Moreover, in \cite{rs1} the extra dimension is compactified by
$S_1/Z_2$, but the size of the compactification scale is not determined
by the dynamics of the model and we need some extra mechanism such as
so-called Goldberger-Wise mechanism \cite{gol}.
On the other hand, in the self-tuning mechanism \cite{ham, kac} we
need not fine-tune on the brane tension. However it is usually
difficult to avoid the appearance of naked singularities in the universe
\cite{csa}. Some warped compactification models in six dimensions
succeeded to avoid naked singularities \cite{hay, kog}.

In this paper we consider a negative tension brane in a five dimensional
bulk with a free scalar field. The cosmological constant in the bulk 
is positive.
We show that it is possible to realize the four dimensional Minkowski 
space without fine tuning. 
Moreover there is no naked singularities in the metric.

\section{The Model}
We consider five dimensional space time with positive cosmological
constant. We introduce a free scalar and a brane whose tension is
$\sigma$. The action is
\begin{eqnarray}
 S&=& \int d^5x \sqrt{-g}(\frac{1}{2}R-\Lambda -\del\phi \cdot \del\phi)
  +\int d^5x \delta(r) \sqrt{-g_4}(-\sigma), \label{ac}
\end{eqnarray}
where $\Lambda$ is positive. This set up in absence of a brane has been
considered in paper \cite{col} and the authors discussed an oscillating
metric.

\begin{figure}[htbp]
 \centerline{\epsfxsize=80mm\epsfbox{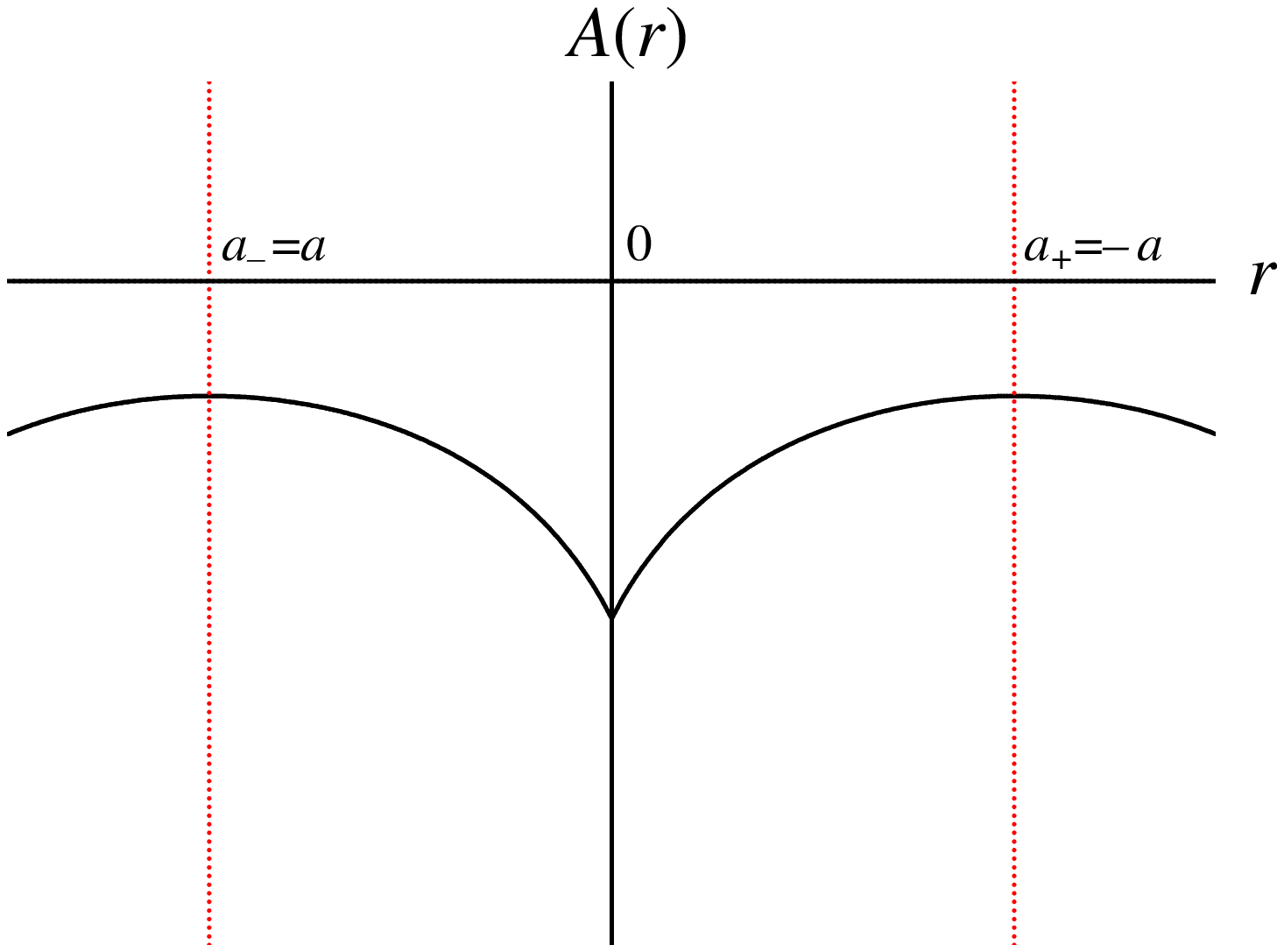}
 \epsfxsize=80mm\epsfbox{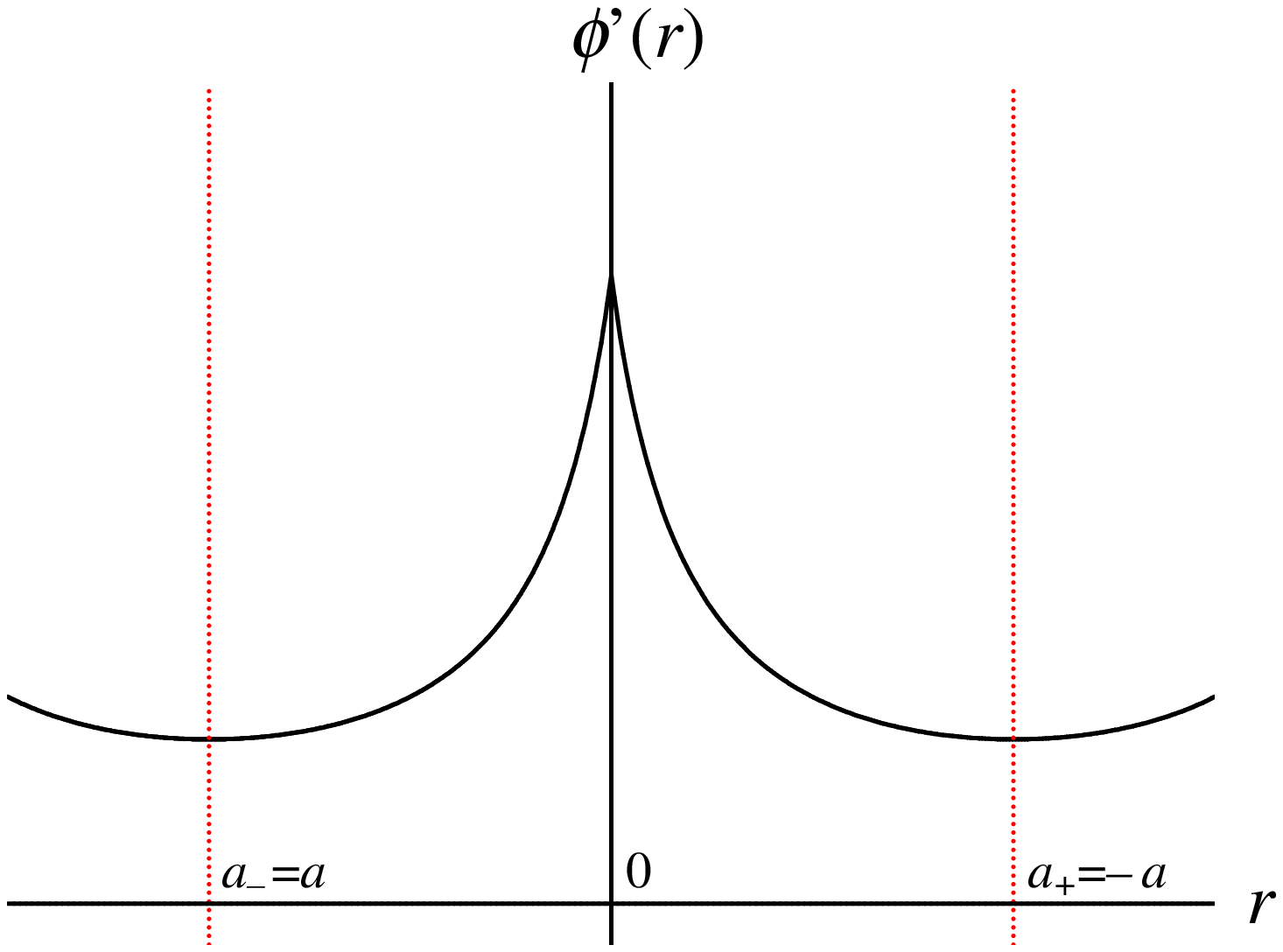}}
 \caption{ Figures for $A(r)$ and $\phi'(r)$. $r$ direction is
 compactified by
 $S_1/Z_2 \sim R/(Z_2 \times Z_2)$. $Z_2$ symmetries are imposed at $r=a$
 and at $r=0$. $A(r)$ and $\phi'(r)$ are even under these $Z_2$ actions.
 The brane sits at $r=0$ and its image points.}
 \label{fig}
\end{figure}

We assume the metric has the following form,
\begin{eqnarray}
 ds^2 &=& e^{2 A(r)}\eta_{\mu\nu}dx^{\mu}dx^{\nu} +dr^2,
\end{eqnarray}
and $\phi$ depends on $r$ only. Then equations of motion are
\begin{eqnarray}
 (\sqrt{-g}\phi')' &=& 0,
  \\
 A''&=&-\frac{2}{3}\phi'^2 -\frac{2}{3}\sigma\delta(r), \label{del}
  \\
 A'^2&=& \frac{1}{6}\phi'^2-\frac{1}{3}\Lambda,
\end{eqnarray}
where $'$ denotes the derivative of $r$. The solution in the bulk is
\begin{eqnarray}
 A(r)&=&\frac{1}{4}\ln [\frac{c_{\pm}}{\sqrt{2\Lambda}} 
   \cos(-4\sqrt{\frac{\Lambda}{3}} (r-a_{\pm}))],  \label{so1}\\
 \phi'(r) &=& c_{\pm}\sqrt{2\Lambda}e^{-4A(r)}.\label{so2}
\end{eqnarray}
There are integration parameters $a_{\pm}$ and $c_{\pm}$,
where $\pm$ labels independent constants in two regions $r<0$ and 
$r>0$ for each.
Those parameters should be partly determined by junction conditions at $r=0$. 
The junction condition for $A$ can be obtained from eq.(\ref{del}).
\begin{eqnarray}
   A'(+ \epsilon) - A'(- \epsilon) = - \frac{2}{3}\sigma. \label{junA}
\end{eqnarray}
There is still another condition for $\phi$, which says $\phi'(0)$
must be continuous.
We can see these conditions allow only an $Z_2$ symmetric solution
with the fixed point $r=0$; $a_{-}=-a_{+}=a$, and $c_{-}=c_{+}=c$.
Without loss of generality we can set $a <0$. Eq.(\ref{junA}) leads to
\begin{eqnarray}
 A'(r)|_{-\epsilon}=\sqrt{\frac{\Lambda}{3}}
  \tan(4\sqrt{\frac{\Lambda}{3}}a)
  &=& \frac{\sigma}{3}, \hs{3ex}
  (-\frac{\pi}{2} <4\sqrt{\frac{\Lambda}{3}}a<0).
  \label{8}
\end{eqnarray}
It is easy to see from above that $\sigma$ must be negative.
The bulk solution has another $Z_2$ symmetry at $r=\pm a$,
so now we identify both points $r=a$ and $r=-a$.
Here we get the extra dimension compactified as $S_1 / Z_2$.
Thus for any negative brane tension the above condition can be satisfied
and $a$ is determined. Therefore we have obtained the regular solution
(see figure \ref{fig})
which realizes a four dimensional Minkowski space without any
fine tuning.

Now we perform naive order analysis. When the absolute value of
 brane tension is small or
equal to $\Lambda$, we take this in the following analysis, from
eq. (\ref{8}) we can obtain rough relation
\begin{eqnarray}
  a \sim \sigma/\Lambda M_*^3,
\end{eqnarray}
where $1/M_*^3$ is the five dimensional Newton constant.
The four dimensional Planck constant $M_p$ is obtained by
the integration of fifth dimension in the action (\ref{ac}) as follows,
\begin{eqnarray}
 M_p^2 \sim M_*^3 |a| \sim \sigma/\Lambda .
\end{eqnarray}
Thus when we take
\begin{eqnarray}
 \Lambda^{1/2} \sim M_* \sim |\sigma|^{1/4} \sim 10^{18} {\rm GeV},
  \label{ton}
\end{eqnarray}
the Plank constant can be correctly obtained as $M_p\sim 10^{18}$
GeV. So every physics is similar to the flat $S_1/Z_2$ compactification.

\section{Discussion}

We have constructed a simple brane model in five dimensions with a
positive bulk cosmological constant. The fifth direction is compactified by
$S_1/Z_2$ and its radius is determined by the brane tension which must
be negative. However we do not need to relate the brane
tension with five dimensional cosmological constant; no fine tuning is
needed. Then it gives one solution for the cosmological constant
problem.

The fixed radius of the extra dimension implies 
that our model has no massless radion mode.
Five dimensional brane world senario has fifth dimensional diffeomorphism 
$r \to r + \xi(x^{\mu},r)$ in the bulk. 
With two branes, just like in Randall-Sundrum (RS) model 
\cite{rs1}, $\xi(x^\mu,r)$ 
is not just a gauge parameter anymore,
but includes a physical mode; a distortion of relative position
between the two branes.
In RS model, this mode is massless and called radion.
It is a reflection that the size of the proper distance between 
the two branes is a modulus in RS model. 
On the other hand in our model, the background figure \ref{fig} manifestly
has no such modulus, and we can conclude there is no massless radion 
in our model.

One comment is in order; in the self-tuning mechanism
\cite{ham, kac, csa} flat four
dimensional space-time can be realized whatever the brane tension
is. In such scenaio, there is a serious problem that
it is difficult to avoid the appearance of singularity/ies 
within a finite length from the brane.
However, our model does not suffer from this problem {\it i.e.}
the metric is regular everywhere
because there is a brane in the middle of the bulk 
before reaching a would-be singularity, 
and the brane now becomes the fixed point of $Z_2$ orbifold.

It is interesting to investigate perturbations of our model from many
point of view, especially to confirm the above naive discussion about 
a radion mode, or to discuss the stability of our model.
Although this investigation seems complicated, it's being achieved elsewhere.

\vs{5mm}

After finishing this work we have been informed of the paper by Jihn E. Kim, 
Bumseok Kyae and Hyun Min Lee \cite{Kim:2001ez}(accepted in NPB).
There they discussed the `dual' model, 
i.e. they considered a 3-form gauge field
which is Poincare dual to a scalar field neglecting boundary 
conditions.
They further introduced two 3-branes to get the compact extra dimension.

\vs{10mm}

\noindent
{\Large\bf Acknowledgment}

T.H would like to thank N.\ Okada and P. Berglund for valuable discussion.
T.H and R.K was supported in part by JSPS Research Fellowships for Young
Scientists. S.H and T.H are grateful also to the organizers of
YITP-workshop (10-13 in July, Kyoto in Japan), where a part of this work
was carried out.


\end{document}